\documentclass[aps,prl,twocolumn,amsmath,amssymb,longbibliography,superscriptaddress]{revtex4-2}
\usepackage{graphicx}
\usepackage{color}
\usepackage[hidelinks]{hyperref}
\usepackage{orcidlink}
\usepackage{amsmath}
\usepackage{amsthm}
\usepackage{physics}
\usepackage{amsfonts}
\usepackage{times,txfonts}
\hypersetup{colorlinks=true, linkcolor=magenta, citecolor=magenta, urlcolor=magenta}

\begin{document}
\author{George Mihailescu\,\orcidlink{0000-0002-0048-9622}}
\email[]{george.mihailescu96@gmail.com}
\affiliation{School of Physics, University College Dublin, Belfield, Dublin 4, Ireland}
\affiliation{Centre for Quantum Engineering, Science, and Technology, University College Dublin, Dublin 4, Ireland}

\author{Pablo M. Poggi\,\orcidlink{0000-0002-9035-3090}}
\affiliation{Department of Physics, SUPA and University of Strathclyde, Glasgow G4 0NG, United Kingdom}

\title{Spectrally local geometric response at the onset of many-body quantum chaos}

\begin{abstract}
We introduce the spectral density of geometric response, a measure of eigenstate sensitivity across an energy spectrum. Applied to many-body quantum systems, it reveals an exponentially sensitive integrability-to-chaos crossover, where eigenbasis deformations first accumulate in localized spectral regions before spreading throughout the spectrum. Physically motivated random-matrix ensembles reproduce this behaviour, whereas Gaussian ensembles do not, indicating universal features of the route to chaos that are absent from featureless random-matrix models.

\end{abstract}
\date{\today}
\maketitle

\emph{Introduction.---}The spectral decomposition of a parameter-dependent Hamiltonian \(\hat{H}(\lambda)\) provides two fundamentally intertwined structures: the energetic landscape encoded by its eigenvalues \(E_n\!\equiv\!E_n(\lambda)\) and a geometric manifold defined by the parameter dependence of its eigenstates \(\vert n\rangle\!\equiv\!\vert n (\lambda)\rangle\). Spectral properties have long provided indispensable diagnostics of many-body phenomena, with level statistics, spectral correlations, and gap scaling revealing signatures of quantum chaos and criticality~\cite{PhysRevB.75.155111,PhysRevLett.52.1,PhysRevX.8.021062,PFEUTY197079TFIM,sachdev1999quantum}. Quantum geometry has emerged as a highly expressive language for quantifying how eigenstates deform under changes of Hamiltonian parameters~\cite{Provost1980,PhysRevLett.99.100603,PhysRevLett.99.095701}. Components of the quantum geometric tensor---and closely related quantities such as fidelity susceptibility and quantum Fisher information---serve as sensitive diagnostics of quantum criticality and multipartite entanglement, while also acquiring operational significance in the design of quantum sensors and optimal control strategies~\cite{PhysRevA.78.042105,pnas.1619826114,gu2008fidelity,j8c7-v2hd,balducci2026nontraversablequantumphasetransitions,zoller,PhysRevLett.105.120501,PhysRevE.76.022101}. 

Despite their inherent physical connection, spectral and geometric perspectives are commonly treated separately. We bridge this divide by introducing the \emph{spectral density of geometric response}, an energy-resolved distribution that assigns each eigenstate a weight determined by its geometric sensitivity. As a first application, we use this framework to study quantum many-body chaos. For two paradigmatic spin models, we show that the entropy of the spectral density of geometric response witnesses a reorganization of geometric sensitivity across the integrability to chaos crossover. Specifically, our results indicate that chaos initially develops through spectrally \emph{local} concentrations of geometric sensitivity, becoming maximally localized at a scale inversely proportional to system dimension. The proposed approach provides an exponentially sensitive probe for integrability breaking (in the spirit of \cite{Pandey_2020}), while concurrently providing fine-grained information about how chaos develops in the spectrum. 
Extending this analysis to random-matrix models, we find a similar intermediate structure emerges when the ensemble retains a physically motivated structure. Conventional Gaussian ensembles display qualitatively different behavior, suggesting that while the integrable and chaotic limits are well-modelled by featureless random matrices, the route to chaos may display more subtle, yet still universal, features.

\begin{figure}[t]
    \centering
    \includegraphics[width=1\linewidth]{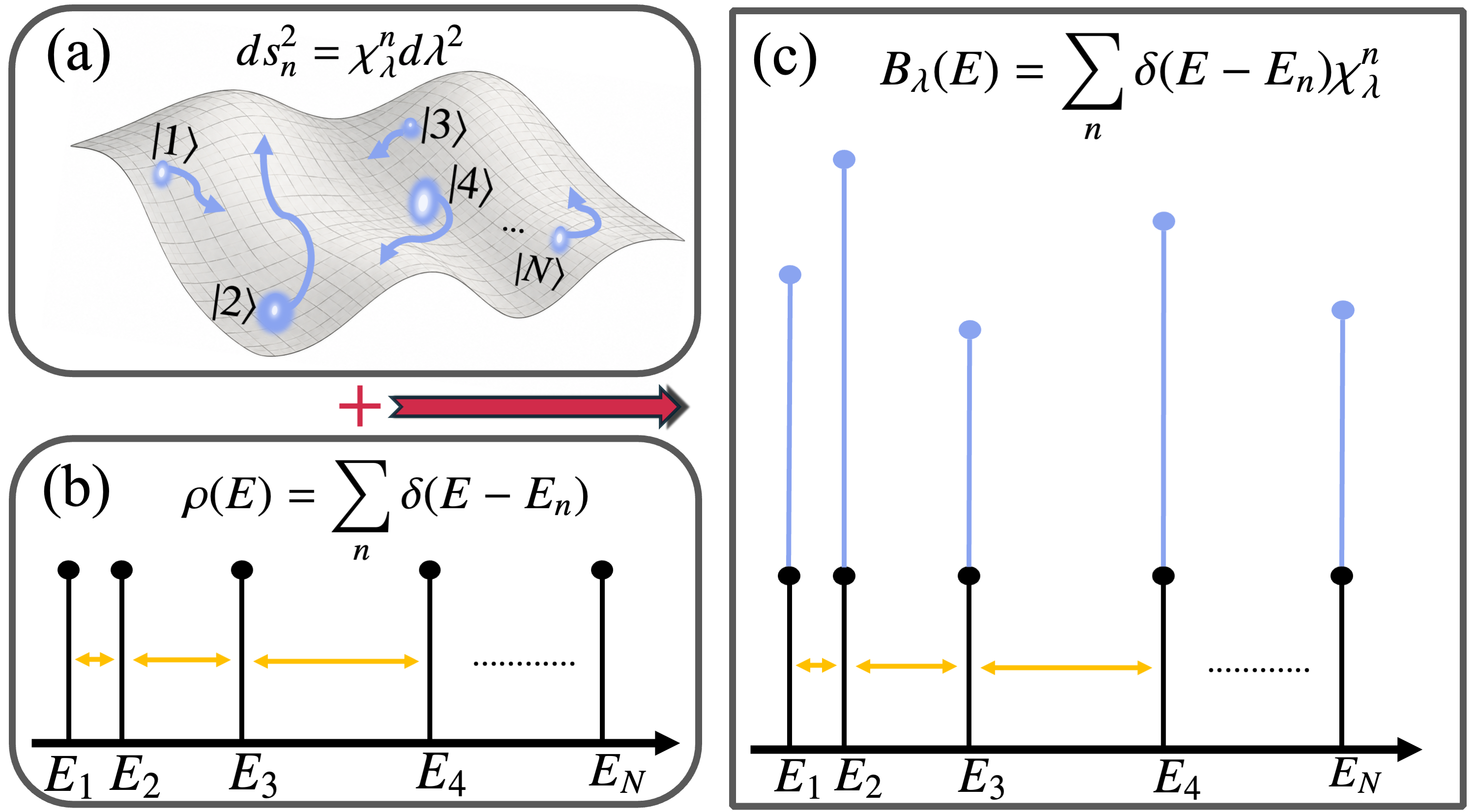}
    \caption{\emph{Schematic}: Spectral density of geometric response. (a) Fidelity susceptibility \(\chi_\lambda^n\) quantifies infinitesimal geometric displacements of eigenstate \(\vert n \rangle\) in response to perturbation \(d\lambda\). (b) Discrete density of states \(\rho(E) = \sum_n\delta(E-E_n)\) locates each eigenstate at energy \(E_n\). (c) Weighting each spectral line by its corresponding \(\chi_\lambda^n\) yields \(B_\lambda(E)\), resolving geometric sensitivity in the spectrum.}
    \label{fig:SDGR_depict}
\end{figure}

\emph{Formalism.---}Given a parameter-dependent Hamiltonian \(\hat{H}(\lambda)\), we consider an energy-resolved distribution of fidelity susceptibility,
\begin{equation}
    \label{eq:fid_suc_dos}
    B_\lambda(E) = \sum_n \delta \left( E - E_n\right) \chi_\lambda^n \;,
\end{equation}
where \( \chi_\lambda^n  = \langle \partial_\lambda n \vert \partial_\lambda n \rangle - \langle \partial_\lambda n \vert n \rangle \langle n \vert \partial_\lambda n \rangle \) is the fidelity susceptibility of eigenstate \( \vert n \left( \lambda \right) \rangle = \vert n \rangle \) with respect to the parameter \( \lambda \)~\cite{Provost1980,PhysRevLett.99.095701}. The quantity \( B_\lambda(E)\) can be interpreted as a spectral density of geometric response (SDGR). Similar to the density of states, \( \rho(E) = \sum_n \delta \left( E - E_n \right)\), the SDGR resolves the spectrum in energy. Assigning the fidelity susceptibility, \( \chi_\lambda^n \), as a weight to each eigenstate, Eq.~\eqref{eq:fid_suc_dos} describes which parts of the eigenbasis are most sensitive to infinitesimal variations in \( \lambda \). The SDGR therefore provides a unified representation of both the spectral and geometric response of a system.

The total spectral weight of the SDGR is fixed by the norm of the adiabatic gauge potential (AGP). Explicitly,
\begin{equation}
    \label{eq:agp_sdgr_relation}
    \frac{1}{\mathcal{D}} \int dE\: B_\lambda \left( E \right) = \frac{1}{\mathcal{D}} \sum_n \chi_\lambda^n \equiv \| \mathcal{A}_\lambda \|^2\;,
\end{equation}
where \( \mathcal{D} \) is the Hilbert-space dimension and \( \| \mathcal{A}_\lambda \|^2  \!=\! \mathcal{D}^{-1} \text{Tr} (\mathcal{A_\lambda}^2) \!=\! \mathcal{D}^{-1}\sum_n \chi_\lambda^n\) denotes the normalized Hilbert-Schmidt norm of the AGP. Eq.~\eqref{eq:agp_sdgr_relation} defines a sum rule for the SDGR: the integrated spectral weight is fixed by the global geometric response of the eigenbasis to the perturbation generated by \(\lambda\). Therefore, while the AGP norm, \(\|\mathcal{A}_\lambda\|^2\), measures the total geometric response of the spectrum~\cite{pnas.1619826114,Pandey_2020}, the SDGR identifies the energies at which this response is concentrated, with \(\|\mathcal{A}_\lambda\|^2\) being the zeroth moment of \(B_\lambda\left(E\right)\).

Normalizing \(B_\lambda\left(E\right)\) by its total weight allows us to develop a fundamental understanding of how the system responds across the spectrum to variations in \(\lambda\) by defining the distribution:
\begin{equation}
    \label{eq:fid_suc_dist}
    P_\lambda\left(E\right) = \frac{B_\lambda\left(E\right)}{\int dE B_\lambda\left(E\right)} = \frac{\sum_n \delta \left(E - E_n\right)\chi_\lambda^n}{\sum_n \chi_\lambda^n}\;,
\end{equation}
with probabilities \(p_n = \chi_\lambda^n/\sum_m \chi_\lambda^m\) denoting the geometric response of eigenstate \(|n\rangle\) relative to the total response of the full eigenbasis. Therefore, while the total geometric response is fixed by the AGP norm, \(P_\lambda\left(E\right)\) and its statistical moments describe how that response is distributed throughout the spectrum. 

This distinction is important, because different physical systems may by characterized by the same total geometric weight---as quantified by the AGP norm. Nevertheless, the microscopic origin of that response may differ substantially. For example, in one system, the response may be broadly distributed over a finite fraction of the eigenbasis, whereas in another it may be concentrated in a small number of anomalously sensitive states. To quantify this structure, we consider the Shannon entropy of the SDGR,
\begin{equation}
    \label{eq:sdgr_entropy}
    S_\lambda = - \sum_n p_\lambda^n \log{p_\lambda^n} = - \sum_n\frac{\chi_\lambda^n}{\sum_m \chi_\lambda^m}\log{\left( \frac{\chi_\lambda^n}{\sum_m \chi_\lambda^m}\right)}\;.
\end{equation}
The normalized entropy \(S_\lambda / \log{\mathcal D}\) quantifies the relative spread of the geometric response throughout the spectrum. An entropy close to zero indicates that the total eigenbasis deformation is dominated by a small number of highly sensitive eigenstates, whereas an entropy close to unity indicates that the response is distributed uniformly across the spectrum. The spectral concentration of the geometric response follows from Eq.~\eqref{eq:sdgr_entropy} as \(\Pi_1^\lambda = {\mathcal D}^{-1} e^{S_\lambda}\), measuring the effective fraction of Hilbert space the geometric response occupies. We can further distinguish whether the eigenbasis deformation is shared uniformly among the participating states or dominated by a smaller subset through the Rényi gap
\begin{equation}
    \label{eq:renyi_gap}
    \Delta_{12}^\lambda = \log{(\Pi_1^\lambda / \Pi_2^\lambda)}\;,
\end{equation}
where \(\Pi_2^\lambda = [\mathcal D \sum_n (p_\lambda^n)^2]^{-1}\) is the Rényi-2 participation fraction. Eq.~\eqref{eq:renyi_gap} vanishes when the response is equally weighted among the states whereas large values signal the eigenbasis deformation is localized to specific spectral regions.

The statistical moments of the normalized SDGR characterize where the geometric response is concentrated. The first moment of Eq.~\eqref{eq:fid_suc_dist}
\begin{equation}
    \label{eq:mean_dist}
    \langle E \rangle_\lambda = \int E P_\lambda\left(E\right) dE = \frac{\sum_n E_n \chi_\lambda^n}{\sum_n \chi_\lambda^n}\;,
\end{equation}
identifies the spectral center of mass of the geometric response. A meaningful figure of merit is the dimensionless offset \(\eta_\lambda = \left(\langle E \rangle _\lambda - \overline{E}\right) / \Gamma \) with \(\overline{E} = \text{Tr}(\hat H) / \mathcal D\) the mean spectral energy and \( \Gamma \) the bandwidth of the system. \(\eta_\lambda\) measures the relative distance between where the geometric response is centred relative to the energetic mean of the spectrum in units the spectral width. A value of \(\eta_\lambda = 0\) indicates that the main contribution of the geometric sensitivity originates from the energetic center of mass, while a positive (negative) value implies the eigenbasis sensitivity is biased above (below) the mean energy of the eigenstates. See~\cite{sup_mat} for further details.

\begin{figure}[tb!]
    \centering
    \includegraphics[width=1\linewidth]{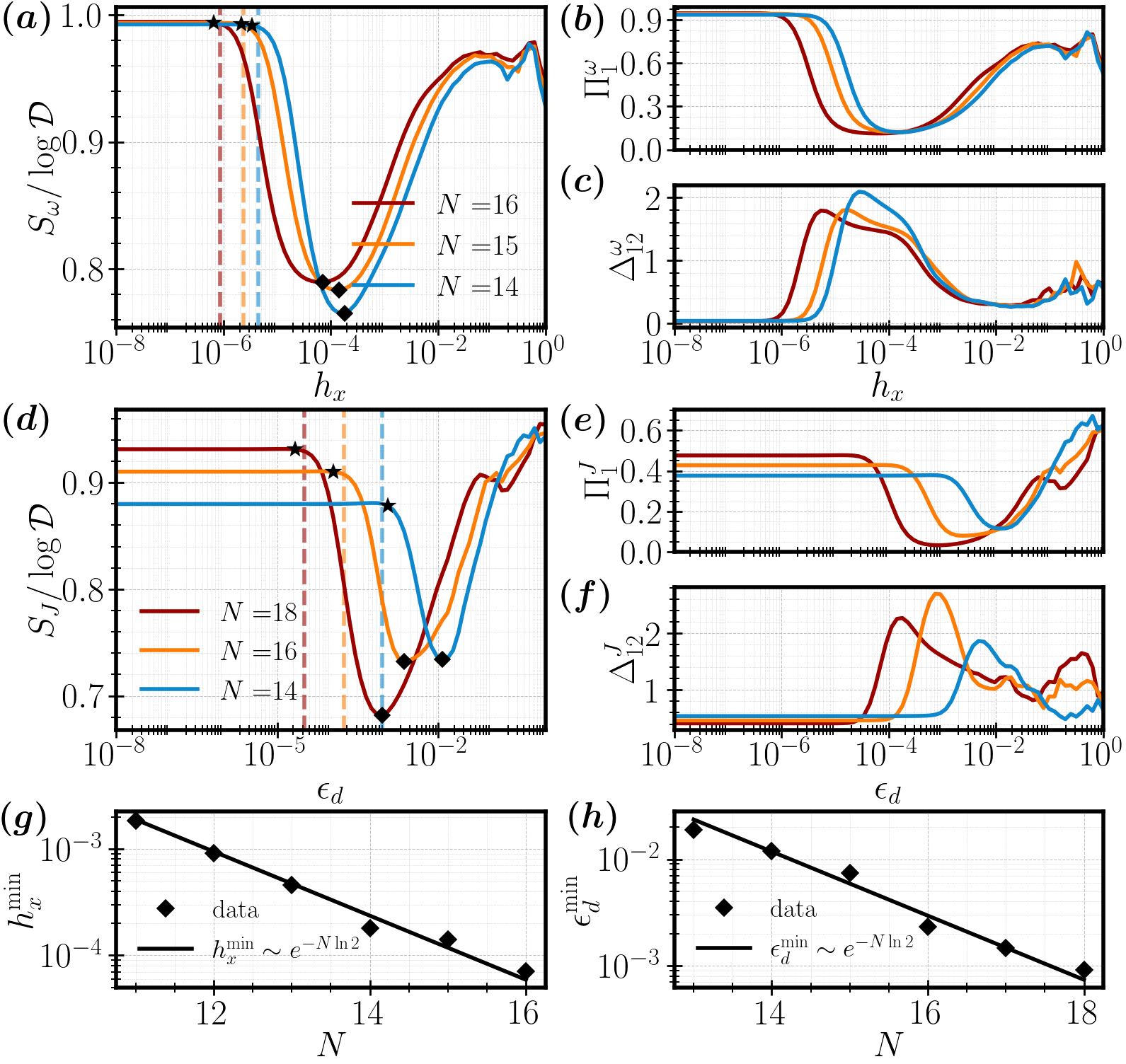}
    \caption{Spectral concentration of geometric response for perturbations in the integrable direction \(X = (\omega,J)\) of the TFIM (a-c) and XXZ (d-f) chains. (a,d) Normalized Shannon entropy \(S_X / \log \mathcal{D}\) as a function of integrability breaking parameters \(Y = (h_x,\epsilon_d)\). Stars mark entropic crossover \(Y^*\) from integrable to chaotic. Vertical dashed lines mark the crossover obtained from AGP norm \(\| \mathcal{A}_X \|^2/N\). Diamonds denote entropic minimum \(Y^{\rm{min}}\). (b,e) Shannon participation fraction, \(\Pi_1^X = e^{S_X}/\mathcal D\), giving the effective fraction of the eigenbasis carrying the response. (c,f) Rényi gap, \(\Delta_{12}^X = \log{(\Pi_1^X/ \Pi_2^X)}\), measuring how unequal the geometric weighting is among the states. (g,h) Entropy minimum position finite-size scaling \(Y^{\text{min}}\!\sim\!e^{-N\log{2}}\). \emph{Parameters.} TFIM: \(\omega \!=\!1\), \(g \!=\! 0.75\). XXZ: \(J \!=\! 1\), \(\Delta \!=\! 1.1\).}
    \label{fig:entropy_results}
\end{figure}

The SDGR admits an equivalent formulation in terms of quantum Fisher information (QFI). For a family of pure states, the fidelity susceptibility is the diagonal component of the quantum metric, and is related to the QFI by \(\mathcal I_\lambda^n \!=\! 4\chi_\lambda^n\)~\cite{caves1994statisticaldistance}. The corresponding QFI spectral density is therefore \(B_\lambda^{\text{QFI}}(E) \!=\! 4 B_\lambda(E)\). This overall factor cancels upon normalization, leaving the probability distribution \(P_\lambda(E)\) [Eq.~\eqref{eq:fid_suc_dist}] unchanged. From a metrological perspective, the SDGR provides a systematic framework for identifying geometric structures across the full spectral manifold underlying emergent many-body phenomena, such as quantum criticality, enabling quantum enhanced sensing~\cite{PhysRevA.78.042105,PhysRevLett.99.100603,PhysRevLett.99.095701,MONTENEGRO20251,quantum_chaotic_sensors,PhysRevE.110.024135,mihailescu_prxq}. 

\emph{Application to quantum chaos.---}The AGP norm is a highly sensitive diagnostic of quantum chaos, scaling polynomially with system size in integrable systems and exponentially in chaotic ones~\cite{Pandey_2020,Kim_2026,92z8-bqx9}. Its crossover detects exponentially weaker integrability breaking perturbations than conventional spectral diagnostics~\cite{PhysRevLett.52.1,PhysRevB.75.155111,PhysRevE.81.036206,ABUELENIN2018564,Dettmann_2017,Dyson_1,PhysRevE.55.4067,PhysRevX.8.021062}. Through its sum rule [Eq.~\eqref{eq:agp_sdgr_relation}], the SDGR retains this diagnostic information, while also revealing features of the integrability-to-chaos crossover inaccessible to the AGP norm alone.

We construct the SDGR by employing the regularization scheme of Ref.~\cite{Pandey_2020}. In particular, we consider the regularized fidelity susceptibility
\begin{equation}
    \label{eq:mu_fid}
    \chi_\lambda^n = \sum_{m \neq n} \frac{\omega^2_{nm}}{\left( \omega^2_{nm} + \mu^2 \right)^2} | \langle m | \partial_\lambda \hat{H} | n \rangle | ^2\;,
\end{equation}
where the fidelity susceptibility in Eq.~\eqref{eq:mu_fid} is written in its spectral representation. Here \(\mu \!=\! N \mathcal{D}^{-1}\) with \(N\) the system size, defines an energy window where energy differences smaller than \(\mu\) suppress the divergence associated with nearly degenerate levels and \(\omega_{mn} \!=\! E_m \!-\! E_n\)~\footnote{We note that the definition of the SDGR does not intrinsically require regularization. We employ the prescription of Ref.~\cite{Pandey_2020} to allow a direct comparison with the AGP-norm crossover.}. We apply our formalism to two spin-chain models in which integrability is broken by alternate mechanisms. We consider a transverse-field Ising chain,
\begin{equation}
    \hat{H}_{\text{TFIM}} = \omega\sum_{i=1}^{N} \hat{\sigma}_z^{(i)} - g \sum_{i=1}^{N-1} \hat{\sigma}_x^{(i+1)} \hat{\sigma}_x^{(i)} - h_x \sum_{i=1}^{N} \hat{\sigma}_x^{(i)}\;,
\end{equation}
in the presence of an integrability breaking longitudinal field \(h_x\). We set \(\omega \!=\! 1\) and \(g \!=\! 0.75\). We also consider an XXZ chain~\cite{PhysRev.112.309,PhysRev.150.321},
\begin{equation}
    \hat{H}_{\text{XXZ}} = J\sum_{i=1}^{N-1} \left( \hat{\sigma}_{x}^{(i+1)}\hat{\sigma}_x^{(i)} + \hat{\sigma}_{y}^{(i+1)}\hat{\sigma}_y^{(i)}  \right) + \Delta \sum_{i=1}^{N-1} \hat{\sigma}_{z}^{(i+1)}\hat{\sigma}_z^{(i)} + \epsilon_d \hat{V}\;,
\end{equation}
with \(\hat V \!=\! \hat{\sigma}_z^{[(N+1)/2]}\) a single site defect and \(\epsilon_d\) characterizes the degree of integrability breaking~\cite{LFSantos_2004}. We set \( J \!=\! 1\) and \(\Delta \!=\! 1.1\). In contrast to the TFIM case, where integrability is broken globally, the XXZ model integrability is broken by a single local defect. These two models therefore provide a complementary test for probing local and global integrability breaking. As standard, we consider open-boundary conditions and focus on a specific symmetry sector for each model. For the XXZ model, we consider the largest fixed-magnetization subspace and for the TFIM, we study the (largest) even spatial-reflection sector. Note \(\mathcal{D}\) therefore represents the dimension of the constructed symmetry sector. In each case, we choose the integrable direction as the perturbation probing the fidelity susceptibility and AGP norm, \(\lambda \!=\! (\omega, J) \!\equiv\! X\), for the TFIM and XXZ respectively. The integrability breaking parameters will be referred to as \(Y \!\equiv\! (h_x,\epsilon_d) \).

\begin{figure}[tb!]
    \centering
    \includegraphics[width=1\linewidth]{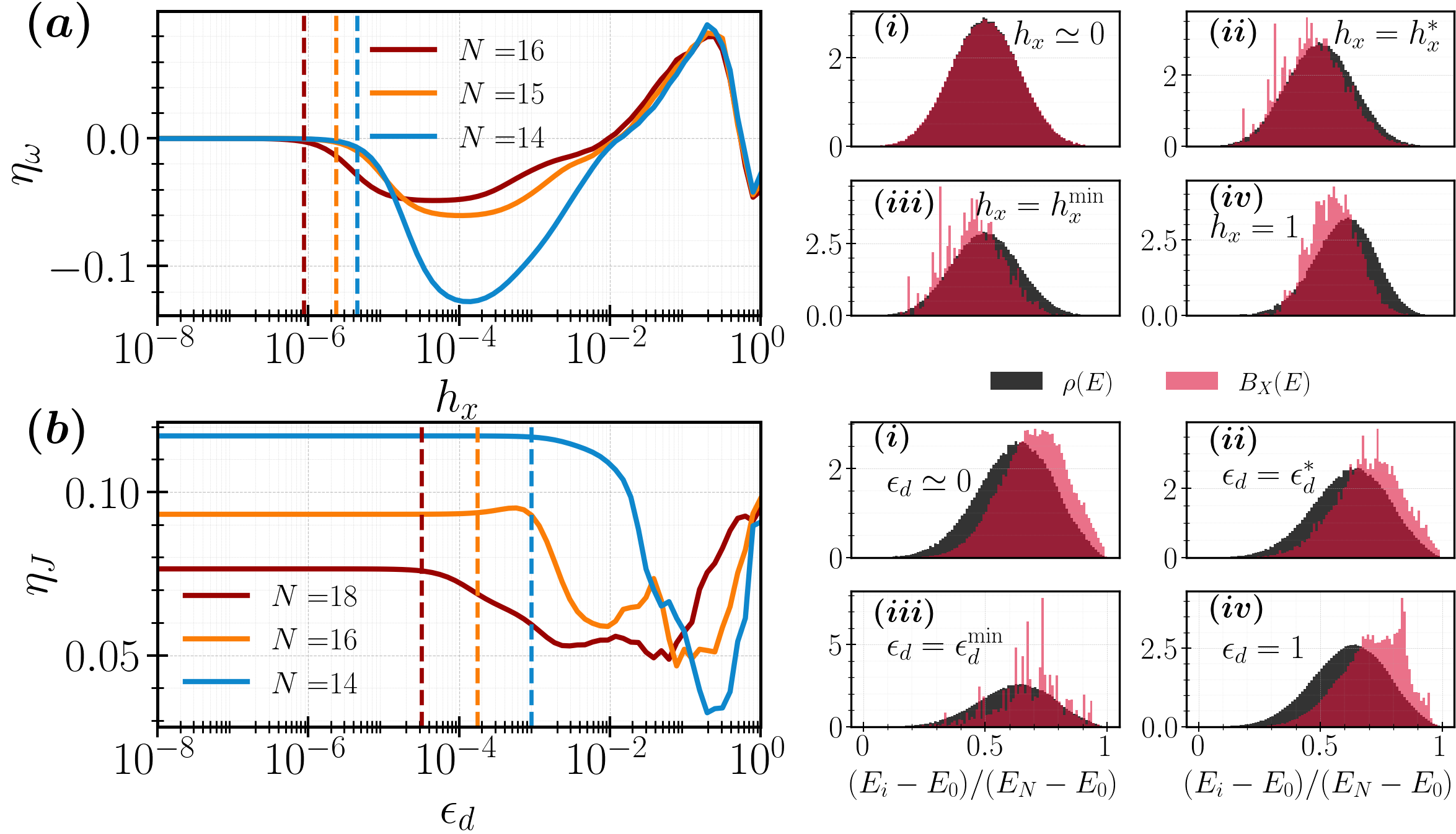}
    \caption{Energy-resolved reorganization of the geometric response corresponding to Fig.~\ref{fig:entropy_results}. (a,b) The dimensionless offset \(\eta_X \!=\!(\langle E \rangle_X - \overline{E}) / \Gamma\) as a function of integrability breaking parameter \(Y \!=\! (h_x,\epsilon_d)\) for the TFIM (a) and XXZ chain (b) respectively. Vertical dashed lines mark the crossover obtained from \(\| \mathcal A_X \|^2/N\). Comparison of the density of states \(\rho(E)\) (dark histogram) with the SDGR \(B_X(E)\) (magenta histogram) at: (i) the integrable plateau \(Y \!\simeq\!0\), (ii) critical perturbation value \(Y \!=\!Y^*\), (iii) entropic minimum point \(Y \!=\!Y^{\text{min}}\), and (iv) strongly chaotic point \(Y\!=\!1\). Shown for \(N\!=\!16\) and \(N\!=\!18\) for the TFIM and XXZ chains.}
    \label{fig:summary_stat}
\end{figure}

Fig.~\ref{fig:entropy_results} shows the normalized Shannon entropy \(S_X / \log{\mathcal{D}}\) of the SDGR as a function of the integrability-breaking perturbation \(Y\) for the TFIM [panel (a)] and the XXZ chain [panel (d)]. In both models, the entropy reveals three distinct regimes. First, for \(Y\!\ll\!Y^*\), it forms an integrable plateau, indicating that the perturbation is too weak to cause a reorganization of the eigenbasis in response to the perturbation. For the TFIM, the plateau saturates near its maximal value \(S_\omega / \log\mathcal D \!\simeq\! 1\) for all system sizes shown. The large participation fraction \(\Pi_1^\omega \!=\! e^{S_\omega}/\mathcal D\) [panel (b)] demonstrates that the response is carried by majority of the eigenbasis, while the vanishing Rényi gap \(\Delta_{12}^\omega\) [panel (c)] shows the states contribute comparable shares of the total eigenbasis deformation. By contrast, the XXZ plateau is system-size dependent, although \(S_J/\log \mathcal D\!\to\!1\) with \(N\). Interestingly \(\Pi_1^J\!\to\!0.5\) [panel (e)] shows the response is carried by a reduced fraction of the eigenbasis. These differences are likely attributable to the different types of integrability of these models (i.e. interacting vs non-interacting), and the different types of perturbations leading to chaos (i.e. spatially local vs global). Nevertheless, the vanishing Rényi gap [panel (f)] indicates that the response is shared relatively evenly within this participating subset, consistent with the TFIM. The spectral center of the response, \(\eta_X\), exhibits a corresponding three-regime structure, remaining constant over the same integrable plateau identified by the entropy [Fig.~\ref{fig:summary_stat}(a,b)].


The second regime describes the crossover from the constant plateau to an entropic minimum, marked by the stars in Fig.~\ref{fig:entropy_results}(a,d). For comparison, the corresponding crossover scales extracted from the AGP norm are shown by the vertical dashed lines~\cite{Pandey_2020}. The crossover scale is characterized by a critical perturbation strength \(Y^* \!=\! (h_x^*,\epsilon_d^*)\) at which integrability breaking becomes sufficiently strong to induce a sudden spectral reorganization of the geometric sensitivity. In this regime, both \(S_X / \log \mathcal D\) and \(\Pi_1^X\) decrease for both models, accompanied by an increase in the Rényi gap \(\Delta_{12}^X\). Therefore, the previously homogeneous spectral response becomes concentrated within a smaller, unequally weighted subset of eigenstates. This behaviour can be seen explicitly in the histograms of the SDGR [Fig.~\ref{fig:summary_stat}(a,b)(ii)], illustrating an enhanced response develops in selected parts of the spectrum. These results suggest that chaotic behavior does not initially emerge homogeneously throughout the eigenbasis, but nucleates through rare, spectrally \emph{local} rearrangements of sensitivity. The crossover scale \(Y^*\) is consistent with that of the AGP norm~\cite{Pandey_2020}, exhibiting exponential scaling with system size. The entropy crossover detects an early redistribution of the normalized spectral weight, providing physical insight into \emph{what} drives the response of the AGP norm and the mechanism underlying its sensitivity. Extracting a precise scaling exponent is technically challenging. We leave this question open for future work. Likewise, \(\eta_X\) provides a complementary witness of the integrability-breaking crossover on a scale consistent with the AGP norm, becoming non-monotonic as the entropy decreases and showing that the emerging concentration of geometric response is accompanied by a redistribution of sensitivity across energy [Fig.~\ref{fig:summary_stat}(a,b)].

Beyond \(Y^*\), the entropy decreases until it reaches an entropic minimum [diamonds in Fig.~\ref{fig:entropy_results}(a,d)]. A scaling analysis of the location of the minimum reveals that 
\begin{equation}
    Y^{\text{min}} \sim e^{-N \log{2}} \sim\mathcal D^{-1} \;,
\end{equation}
with the same leading scaling observed for the TFIM [panel (g)] and XXZ [panel (h)] chains. In this regime, the effective support of the geometric sensitivity contracts substantially to the restricted set of states dominating the response [panel (b) and (e)], with \(\Pi_1^X\) decreasing toward its minimum, indicating that the response is concentrated over a reduced fraction of the eigenbasis. Correspondingly, \(\Delta_{12}^X\) reaches its maximum, revealing a highly heterogeneous distribution of the geometric weight. The histograms of the SDGR [Fig.~\ref{fig:summary_stat}(a,b)(iii)] provide the corresponding energy resolved picture, showing local spectral regions have a high concentration of the geometric weight. The minimum scale \(Y^{\text{min}}\) therefore marks the point at which the geometric response becomes maximally localized over the eigenbasis, with its weight concentrated in a small number of exceptionally sensitive eigenstates occupying restricted spectral regions~\cite{PhysRevE.104.054105,PhysRevB.111.184203}. Interestingly, the scaling of the entropic minimum is consistent with the observations of Ref~\cite{LeBlond_2021,zc68-wj2f}, who argue the onset of chaos is marked by maxima of the typical fidelity susceptibility.

For \(Y\!>\!Y^{\text{min}}\), the integrability breaking perturbation becomes sufficiently strong for resonances to proliferate and overlap, connecting neighbouring regions of the spectrum. This produces a rise in \(S_X / \log{\mathcal D}\) as the deformation begins to spread across a larger portion of the eigenbasis. Similarly, the participation fraction \(\Pi_1^X\) increases, showing that a growing fraction of eigenstates begin to carry the response, while a decrease of the Rényi gap \(\Delta_{12}^X\) shows the response becomes shared more evenly among those states. We interpret this as a leakage of the chaotic behaviour from the selected local spectral regions at \(Y^{\text{min}}\) into neighbouring eigenstates. This picture is supported by the SDGR histograms in Fig.~\ref{fig:summary_stat}(a,b)(iv), where the SDGR broadens around the states where the localized geometric support was concentrated previously at \(Y^{\text{min}}\) [Fig.~\ref{fig:summary_stat}(a,b)(iii)].

\begin{figure}[tb!]
    \centering
    \includegraphics[width=1\linewidth]{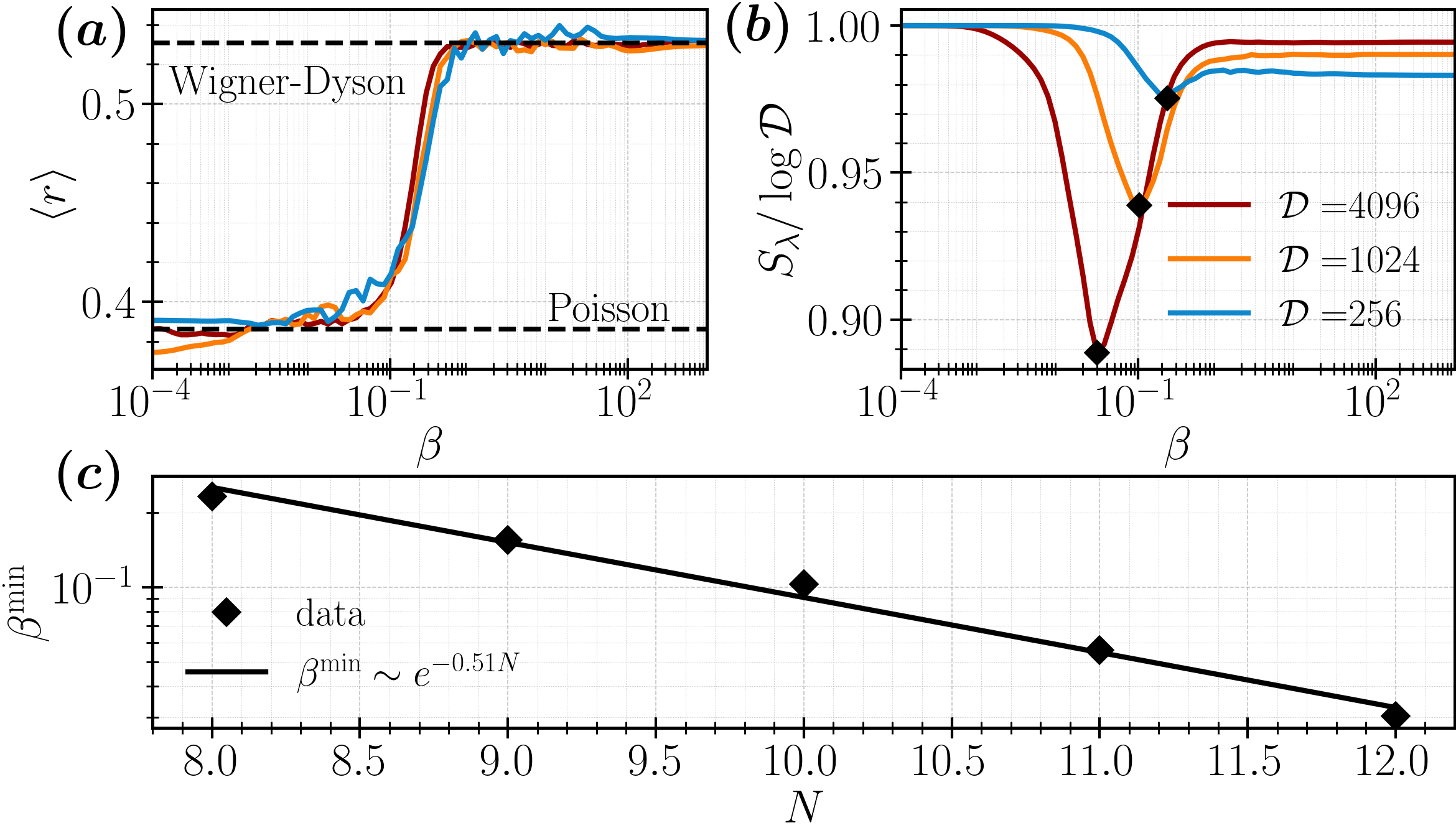}
    \caption{Pauli random matrix theory construction. (a) Level-spacing statistics as a function of integrability breaking parameter \(\beta\). (b) Normalized Shannon entropy \(S_\lambda / \log \mathcal D\) of the SDGR for perturbations in the integrable direction \(\lambda\). Diamonds mark the entropic minimum. (c) Finite-size scaling of the entropy minimum position \(\beta^{\text{min}}\!\sim\!e^{-0.51 N}\). System sizes shown \(N\!=\!8,10,12\) corresponding to \(\mathcal{D}\!=\!2^8,2^{10},2^{12}\).}
    \label{fig:RMT}
\end{figure}

\emph{Random matrix theory.---}Predictions from random matrix theory (RMT) distinguish integrable and chaotic regimes through level-spacing statistics. Integrable many-body systems typically exhibit Poisson statistics, reflecting the absence of level repulsion, whereas chaotic systems develop level repulsion and approach Wigner-Dyson statistics~\cite{7145ab01da7,PhysRevLett.52.1}. However, level-spacing statistics interpolate smoothly between these limits~\cite{rosenzweig1960repulsion,lenz1991}, providing limited insight into the physical process by which chaos emerges. Remarkably, in the physical models considered previously, the SDGR entropy resolves this process as a non-trivial reorganization of the geometric response across the spectrum. Here, we ask whether this structure persists in an RMT setting and if these features are generic to the integrability to chaos transition. 

To address this question, we construct a random matrix model composed solely of one- and two-body Pauli operators \cite{brown2018}, which we term Pauli-RMT~\cite{lenz1991}, defined as
\begin{equation}
    \hat H(\lambda,\beta) = \mathcal{N}(\beta)(\hat H_0 + \lambda \hat H_1 + \beta  \hat V)\;,
    \label{eq:rmt_hami}
\end{equation}
where \( \mathcal N(\beta)^2 \!=\! (1 + \lambda^2_0 + \beta^2)^{-1} \) a normalization factor, with \(\lambda_0\!=\!1\). The operators in Eq.~(\ref{eq:rmt_hami}) are random linear combinations of Pauli strings: \(\hat H_0\) contains single-site \(\hat Z\) terms, \(\hat H_1\) single-site \(\hat X\) and \(\hat Z\) terms, and \(\hat V\) one- and two-site \(\hat X\) and \(\hat Z\) terms. For fixed \(\lambda\), the model is integrable (non-interacting) for \(\beta\!=\!0\) and undergoes a crossover to quantum chaos as \(\beta\) increases, as shown by the mean level-spacing ratio in Fig.~\ref{fig:RMT}(a). We then evaluate \(S_\lambda\) [Eq.~\eqref{eq:sdgr_entropy}], probing the integrable direction of fidelity susceptibility as before, and plot it against the integrability breaking parameter \(\beta\) in Fig.~\ref{fig:RMT}(b). The resulting crossover reveals a hierarchy of scales consistent with the observations of the physical TFIM and XXZ chains. \(S_\lambda / \log \mathcal D\) first departs from its integrable plateau while the mean level-spacing remains essentially Poissonian, exposing eigenstate reorganization before it is visible to conventional spectral statistics. The entropy subsequently reaches a distinct minimum near the chaos transition \(\beta^{\text{min}}\) on the scale at which the level statistics begin to depart from Poisson---marking the point of maximal local concentration of the geometric response. In Fig.~\ref{fig:RMT}(c) we extract the scaling of the entropic minimum and find \(\beta^{\text{min}}\!\sim\!e^{-0.51 N}\) which is different but comparable to the physical models studied. For \(\beta\!>\!\beta^{\text{min}}\), the entropy rises toward a second, chaotic plateau, and the level statistics correspondingly approach their Wigner-Dyson value. The entropy of the SDGR therefore resolves structure \emph{within} the integrability to chaos crossover rather than merely distinguishing its two asymptotic limits. 

Remarkably, when analyzing standard Gaussian RMT ensembles we found no analogous entropic minimum (see \cite{sup_mat}), with the entropy displaying quantitatively distinct features from the physical models and Pauli RMT, despite consistent level-spacing statistics and AGP norms across RMT constructions. This suggests that, while conventional (featureless) random-matrix ensembles capture universal properties of integrable and chaotic limits, the route by which chaos emerges can depend on certain coarse-grained physical properties of the model. The SDGR provides access to the chaotic path taken, offering a means not only to identify the transition toward quantum chaos, but also to probe the physical reorganization of the eigenbasis through which chaos develops. Perhaps surprisingly, the AGP norm remains consistent across the two constructions, but as a global measure of the total geometric response it washes out the spectral redistribution of sensitivity that distinguishes the crossovers. Previously, fidelity susceptibility has been characterized analytically in Gaussian ensembles, providing a universal benchmark for eigenstate sensitivity once random-matrix behavior is established~\cite{Sierant_2019,Berry_2020,PhysRevLett.126.200604}; here, the SDGR resolves how this geometric response reorganizes on the approach to that regime.

\emph{Conclusions.---}Applied to quantum chaos, the SDGR leverages the spectral resolution of geometric deformations throughout the eigenbasis to expose the underlying physical processes responsible for the exponential sensitivity of the AGP norm across the crossover to chaotic behavior. This perspective further uncovers novel insights into the integrable-to-chaotic transition. Given that geometric objects such as the quantum Fisher information can capture universal characteristics of phenomena such as quantum phase transitions~\cite{PhysRevLett.99.100603,PhysRevLett.99.095701,mihailescu_prxq} and non-Markovianity~\cite{fq4l-8v5g,PhysRevA.82.042103,PhysRevA.91.042110,SciPostPhys.15.1.014}, we anticipate that our formulation could unveil novel structures underlying these and related phenomena by exploiting its joint spectral and geometric resolution.

\emph{Acknowledgments.---}G.M. acknowledges support from Taighde \'Eireann - Research Ireland under grant number 24/EPSRC/4121. G.M. would like to thank Phillip Cussen Burke, Shane Dooley, Eoin Carolan, and Eoin O'Connor for fruitful conversations and useful comments. P.M.P. acknowledges support from the Royal Society through a University Research Fellowship (Grant No. URF\textbackslash R1\textbackslash252030). 
\bibliography{bibo}

\newpage
\onecolumngrid 
\subsection{\large{{End Matter}}}
\twocolumngrid

\begin{figure*}[htb!]
    \centering

    \includegraphics[width=0.98\textwidth]{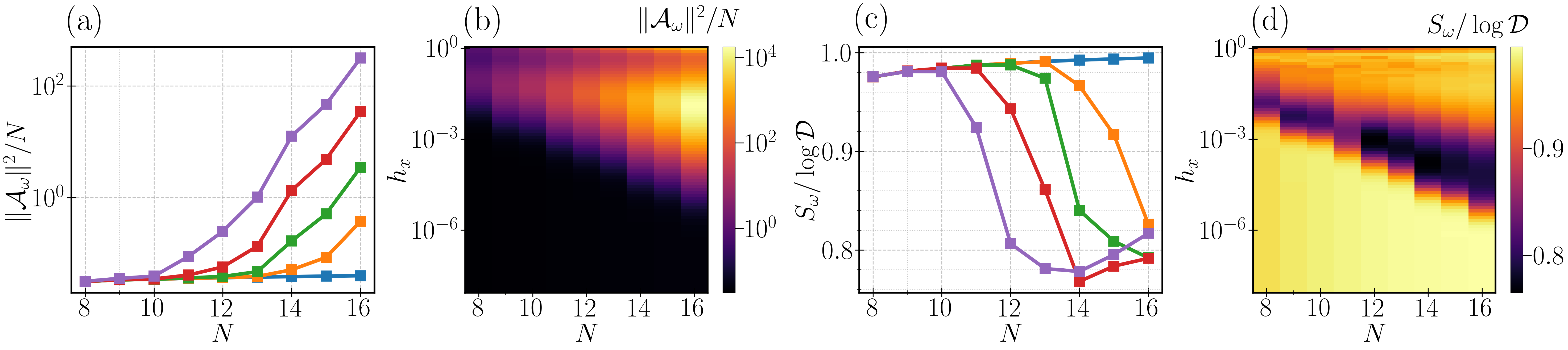}

    \includegraphics[width=0.98\textwidth]{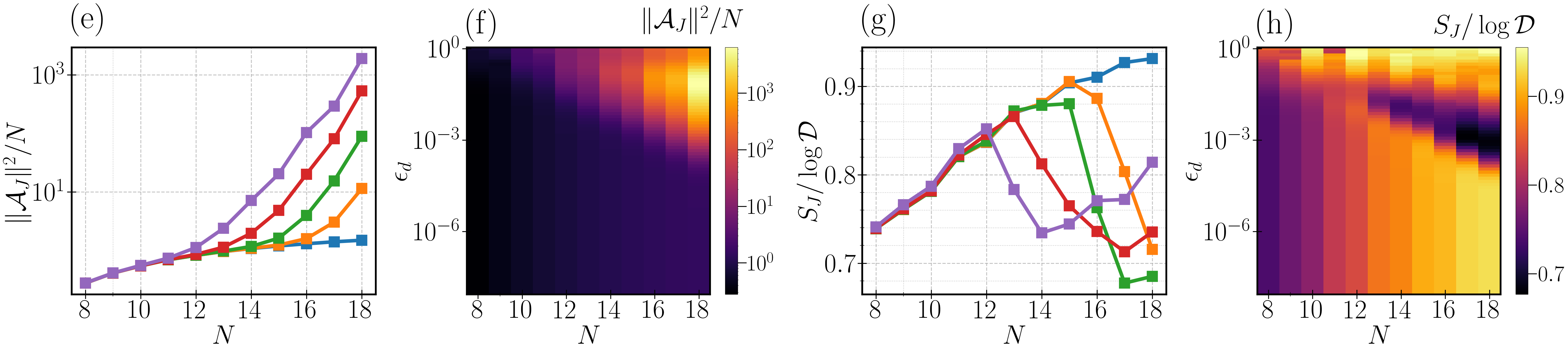}

    \caption{Rescaled AGP norm \(\|\mathcal A_X\|^2/N\) and entropy of SDGR \(S_X / \log{\mathcal D}\) as a function of system size \(N\) for various values of the integrability breaking parameter \(Y\!=\!(h_x,\epsilon_d)\) for the TFIM (a-d) and XXZ (e-h) chains respectively. For the TFIM in panels (a) and (c) the value of integrability breaking parameter chosen is \(h_x\!=\! (10^{-8},1.9\!\times\!10^{-5},3.5\!\times\!10^{-5},1.1\!\times\!10^{-4},3.6\!\times\!10^{-4})\) corresponding to the blue, orange, green, red, and purple lines. For the XXZ in panels (e) and (g) the value of integrability breaking parameter chosen is \(\epsilon_d\!=\!(10^{-8},3.6\!\times\!10^{-4},1.1\!\times\!10^{-3},3.7\!\times\!10^{-3},1.1\!\times\!10^{-2})\) corresponding to the blue, orange, green, red, and purple lines.}
    \label{fig:sm_tfim_xxz_finite_N}
\end{figure*}

\emph{Unified statistical picture.---}We consider the joint empirical distribution
\begin{equation}
    \label{eq:joing_dist}
    \mathcal Q_\lambda(E, \chi) = \mathcal D^{-1} \sum_n \delta(E-E_n) \delta(\chi - \chi_\lambda^n)\;,
\end{equation}
associating to each uniformly sampled eigenstate its spectral position \(E_n\) and corresponding fidelity susceptibility \(\chi^n_\lambda\). Integrating over energy gives the fidelity susceptibility distribution 
\begin{equation}
    q_\lambda(\chi) = \int dE \mathcal Q_\lambda(E,\chi) = \mathcal D^{-1} \sum_n \delta(\chi-\chi_n)\;,
\end{equation}
whose fluctuations and tails have been used to characterize eigenstate sensitivity and chaos~\cite{Sierant_2019,LeBlond_2021,Pandey_2020}. The complementary susceptibility weighted projection of Eq.~\eqref{eq:joing_dist} gives the SDGR \(B_\lambda(E) = \mathcal D \int d\chi \mathcal Q_\lambda(E,\chi) \chi = \sum_n \chi_\lambda^n \delta(E-E_n)\). Operationally, \(q_\lambda(\chi)\) resolves how geometric sensitivity is distributed across eigenstates and \(B_\lambda(E)\) resolves where that sensitivity is concentrated in energy. Together, these distributions provide complementary projections of the same joint spectral-geometric distribution. Despite retaining distinct information, both \(q_\lambda(\chi)\) and \(B_\lambda(E)\) reduce to the same global measure of geometric response: 
\begin{equation}
    \| \mathcal A_\lambda \|^2 = \int q_\lambda(\chi) \chi d\chi = \mathcal D^{-1} \int B_\lambda(E) dE\;.
\end{equation}
From this perspective, the AGP norm admits two equivalent statistical interpretations: (i) the mean fidelity susceptibility of a uniformly sampled eigenstate, and (ii) the total geometric response per eigenstate obtained by integrating the SDGR. However, as a global diagonstic, the AGP norm retains only the overall magnitude of the eigenbasis deformation. The distributions \(q_\lambda(\chi)\) and \(B_\lambda(E)\) resolve complementary information lost in this reduction.

\emph{Alternative interpretation of SDGR.---}The fidelity susceptibility with respect to eigenstate \(\vert n \rangle\) describes the fluctuations (variance) of the adiabatic gauge potential in that respective eigenstate (in the parallel transport gauge):
\begin{equation}
    \chi_\lambda^n = \langle n \vert \mathcal{A}_\lambda^2 \vert n \rangle - \langle n \vert \mathcal A_\lambda \vert n \rangle^2\;.
\end{equation}
From this perspective the SDGR in Eq.~\eqref{eq:fid_suc_dos} can be written as:
\begin{align}
    B_\lambda(E) &= \sum_n \delta(E - E_n)[\langle n \vert \mathcal{A}_\lambda^2 \vert n \rangle - \langle n \vert \mathcal A_\lambda \vert n \rangle^2] \\
    &= \sum_n \delta(E-E_n) [\text{Var}(\mathcal A_\lambda)]_n\\ 
    &= \text{Tr}\left[ \delta(E - \hat H) \mathcal A_\lambda^2 \right]\;,
\end{align}
meaning the SDGR can be interpreted as an energy resolved density of fluctuations of the AGP. Integrating over energy recovers the Hilbert-Schmidt norm of the AGP, i.e., \(\mathcal{D}^{-1}\int dE B_\lambda(E) \!=\! \mathcal D^{-1} \text{Tr}(\mathcal A_\lambda^2) \!=\! \| \mathcal A_\lambda \|^2\) as before. The operational interpretation is that the AGP norm quantifies the total fluctuation strength, while SDGR identifies where in the spectrum these fluctuations are concentrated.

We note that the SDGR admits a natural statistical interpretation. Its total spectral weight can be interpreted as a geometric partition function,
\begin{equation}
    \label{eq:sm_geo_pf}
    \mathcal Z_\lambda^\chi = \int dE B_\lambda(E) = \sum_n \chi_\lambda^n =  \mathcal D \| \mathcal A_\lambda \|^2\;,
\end{equation}
giving the total geometric response of the eigenbasis to the perturbation generated by \(\lambda\). Alternatively, one can interpret it as total fluctuations of the AGP within the eigenbasis. The normalized weight of each eigenstate \(\vert n \rangle\) is then given by \(p_\lambda^n \!=\! \chi_\lambda^n/\sum_m \chi_\lambda^m \!=\! [\mathcal Z_\lambda^\chi]^{-1}\chi_\lambda^n\), measuring the fraction of total geometric response carried by each eigenstate. The normalized SDGR can therefore be written as
\begin{equation}
    P_\lambda(E) = [\mathcal Z _\lambda^\chi]^{-1} B_\lambda(E) = \sum_n p_\lambda^n \delta(E - E_n)\;,
\end{equation}
describing the energy distribution induced by sampling eigenstates according to their geometric sensitivity. In this formulation, the AGP norm fixes the total response, while \(P_\lambda(E)\) describes how that response is distributed throughout the spectrum. Written in this way, the entropy of this geometric ensemble can take the form
\begin{equation}
    \label{eq:em_slam}
    S_\lambda = \log \mathcal Z_\lambda^\chi - \langle \log \chi_\lambda^n \rangle_{p_\lambda}\;,
\end{equation}
where \(\langle \log \chi_\lambda^n \rangle_{p_\lambda} \!=\! \sum_n p_\lambda^n \log \chi_\lambda^n\) is the typical logarithmic fidelity susceptibility. This gives the entropy the operational interpretation as sampling the typical response (or fluctuation of the AGP) compared to the total geometric response of the eigenbasis.

\emph{Finite-size entropy behaviour.---}In the main text we characterize the integrability to chaos crossover by varying the integrability breaking parameter \(Y\!=\!(h_x,\epsilon_d)\) at a fixed system size. Here, we present the complementary finite-size perspective, fixing the perturbation \(Y\) and increasing the system size \(N\). Fig.~\ref{fig:sm_tfim_xxz_finite_N} compares the rescaled AGP norm \(\|\mathcal A\|^2/N\), with the normalized entropy of the SDGR \(S_X/\log \mathcal D\) for the TFIM and XXZ chains, making explicit how the reorganization of the geometric response identified in Fig.~\ref{fig:entropy_results} of the main text develops with increasing system size.

At sufficiently weak integrability breaking, the AGP norm initially follows its integrable finite size behavior--displaying polynomial growth, shown in Fig.~\ref{fig:sm_tfim_xxz_finite_N}(a,e) [see for example the blue line]. As the system size \(N\) increases, each finite non-zero value of \(Y\) eventually reaches a characteristic system size \(N^*(Y)\) beyond which the AGP norm begins to develop exponential growth associated with chaotic eigenstate sensitivity~\cite{Pandey_2020}. Larger values of integrability breaking parameter can reach this crossover at smaller \(N\), whereas weaker perturbations require larger system sizes. This behaviour is seen evidently in the heat maps of Fig.~\ref{fig:sm_tfim_xxz_finite_N}(b,f), where the crossover forms a sharp feature where the value of \(Y\) required to detect integrability breaking decreases as the system size \(N\) increases. This is the fixed-\(Y\) counterpart of the exponentially decreasing crossover scale \(Y^*(N)\) discussed in the main text. 

The entropy provides a microscopic interpretation of the growth of the AGP norm. Recall, the total geometric weight satisfies \(\mathcal Z_X^\chi \!=\! \sum_n \chi_X^n \!=\! \mathcal D \| \mathcal A_X \|^2\), here taken in the integrable direction \(X\!=\!(\omega,J)\) for the TFIM and XXZ chains. Utilizing Eq.~\eqref{eq:em_slam} and defining the geometrically weighted typical fidelity susceptibility as 
\begin{equation}
    [\chi_{X}^{(p)}]_{typ} = \exp{[\langle \log \chi_X^n \rangle_{p_X}]}\;,
\end{equation}
it is easy to show that 
\begin{equation}
    \|\mathcal A_X\|^2 = \Pi_1^X [\chi_{X}^{(p)}]_{typ}\;,
\end{equation}
with \(\Pi_1^X\!=\! e^{S_X}/\mathcal D\) the Shannon participation fraction. Therefore, the rise of the AGP norm contains two distinct pieces of information: the fraction of the eigenbasis carrying the response and the characteristic sensitivity of those participating eigenstates. The decomposition is directly visible in Fig.~\ref{fig:sm_tfim_xxz_finite_N}(c,g). For the TFIM, the departure of \(\|\mathcal A_X\|^2/N\) from its integrable behavior coincides with the sharp decrease of \(S_X /\log \mathcal D\). In the XXZ chain, the integrable entropy has a stronger finite size dependence, but the same phenomenon appears as a downward departure from the integrable curve. Thus, precisely when the total geometric response begins its exponential growth, our formalism reveals the total eigenbasis deformation becomes concentrated over a smaller fraction of the eigenbasis. Furthermore, as \(\Pi_1^X\) decreases while \(\|\mathcal A_X \|^2\) increases, the growth of the norm must initially be driven by a rapid enhancement of the susceptibility of a restricted subset of exponentially sensitive eigenstates. The AGP norm crossover therefore reflects the nucleation of anomalously large eigenstate deformations in rare spectral regions, consistent with the energy resolved pictures presented in Figs.~\ref{fig:entropy_results} and \ref{fig:summary_stat} of the main text. 

Increasing \(N\) at a fixed \(Y\), the entropy passes through its minimum, and then begins to increase. The non-monotonic finite size dependence is the complementary analogue of the entropy minimum \(Y^{\text{min}}(N)\!\sim\!\mathcal D^{-1}\) identified in the main text. Because this scale shifts toward exponentially smaller perturbations as the Hilbert space grows, a fixed value of \(Y\) successively crosses the integrable regime, the region of maximal spectral concentration, and eventually the chaotic behavior proliferates to the rest of the states. Beyond the minimum, the increase of \(S_X / \log \mathcal D\) shows that the geometric response is carried by a growing fraction of the eigenbasis for a fixed perturbation strength. This is despite the AGP norm continuing to grow exponentially with system size. This is because in this regime, the initially localized response has proliferated through resonant mixing into neighbouring spectral regions. Fig.~\ref{fig:sm_tfim_xxz_finite_N} therefore resolves the physical process underlying the AGP norm crossover: the exponential enhancement of the total geometric response is initiated by rare, highly sensitive eigenstates and is subsequently converted into a broadly distributed chaotic response as the system size increases.  







\clearpage
\onecolumngrid


\setcounter{page}{1}
\renewcommand{\thepage}{S\arabic{page}}

\setcounter{equation}{0}
\renewcommand{\theequation}{S.\arabic{equation}}
\renewcommand{\theHequation}{supp.equation.\arabic{equation}}

\setcounter{figure}{0}
\renewcommand{\thefigure}{S\arabic{figure}}
\renewcommand{\theHfigure}{supp.figure.\arabic{figure}}

\setcounter{table}{0}
\renewcommand{\thetable}{S\arabic{table}}
\renewcommand{\theHtable}{supp.table.\arabic{table}}

\setcounter{section}{0}
\setcounter{subsection}{0}
\setcounter{secnumdepth}{2}

\renewcommand{\thesection}{S.\Roman{section}}
\renewcommand{\thesubsection}{\thesection.\Alph{subsection}}

\renewcommand{\theHsection}{supp.section.\arabic{section}}
\renewcommand{\theHsubsection}{supp.subsection.\arabic{section}.\arabic{subsection}}
\subsection*{\large{{Supplemental Material for\\
 ``Spectrally local geometric response at the onset of many-body quantum chaos''}}}

In this Supplemental Material we provide supporting information and data.

\begin{itemize}
    \item In Section~\ref{sm_other_stat_momo} we discuss other statistical moments of the SDGR probability distribution.
    \item In Section~\ref{sm_finite_temp_ext} we extend our formalism to finite temperature.
    \item In Section~\ref{sm_linear_reponse} we connect the SDGR to linear response theory.
    \item In Section~\ref{sm_rmt} we provide further details on the random matrix theory construction. 
\end{itemize}

\section{Other statistical moments of the SDGR}\label{sm_other_stat_momo}

\begin{figure*}[!htb]
\centering
    \includegraphics[width=0.49\textwidth]{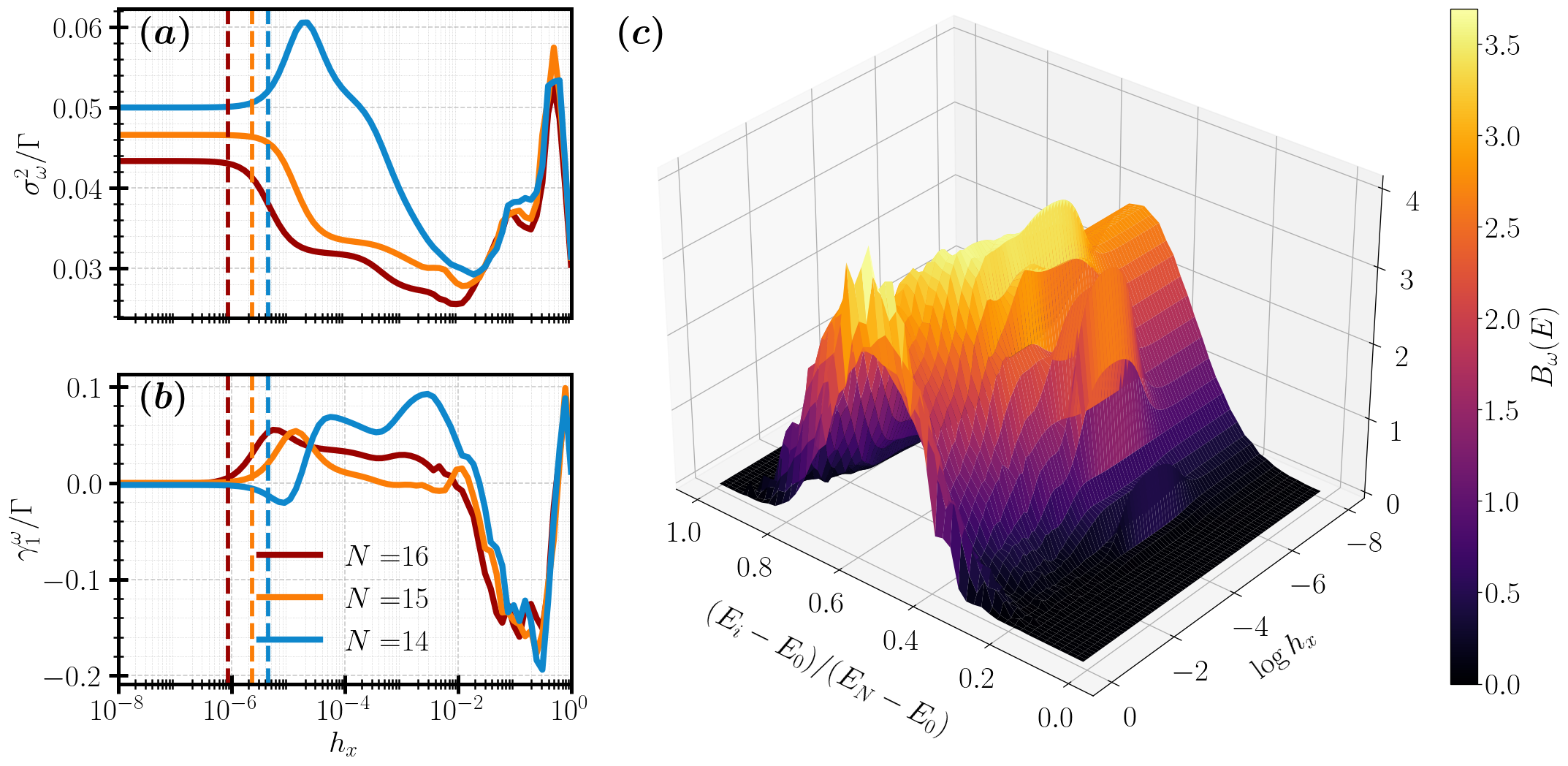}
    \hfill
    \includegraphics[width=0.49\textwidth]{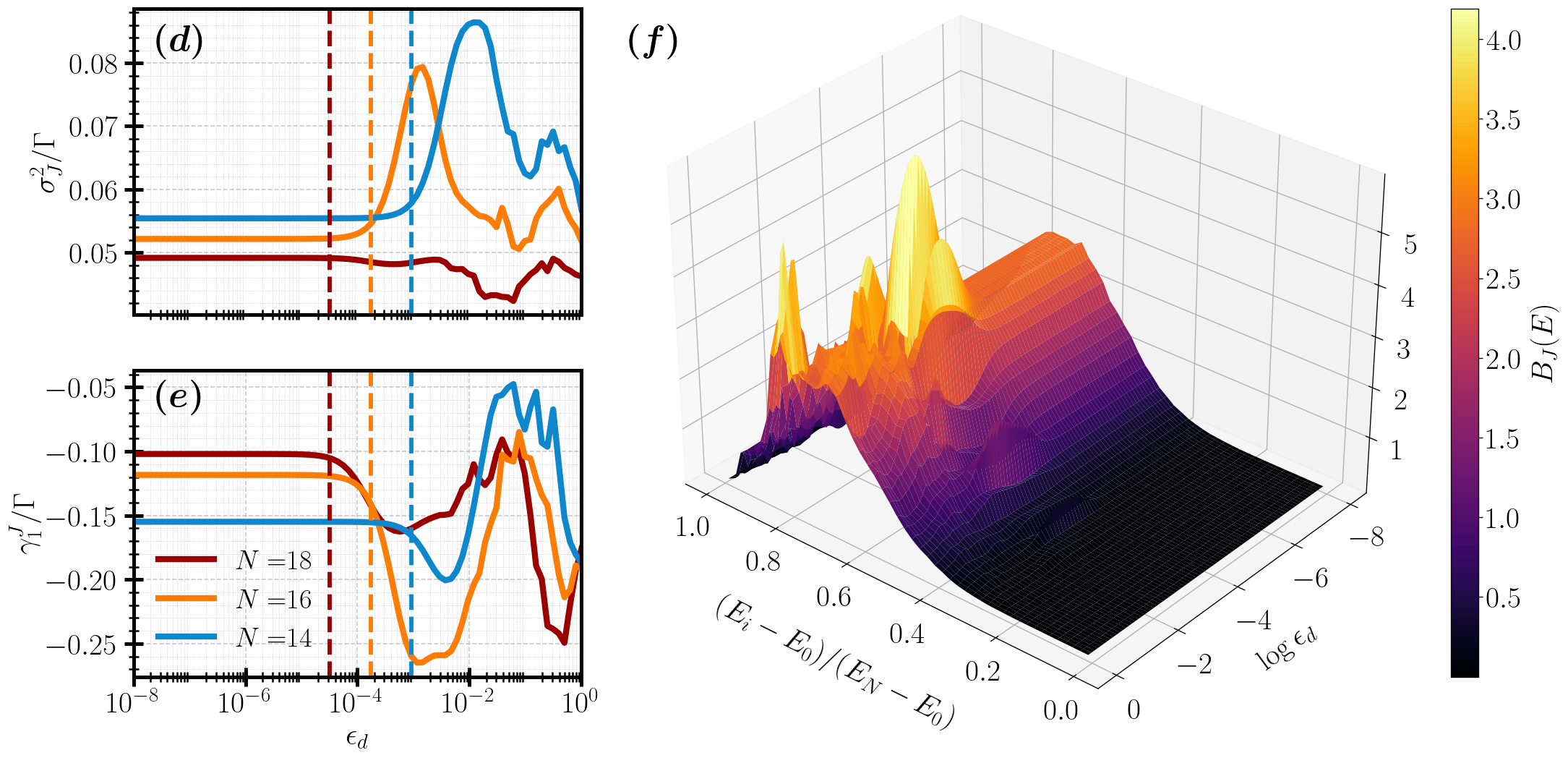}
	\caption{Higher order statistical moments and continuous energy resolved evolution of the SDGR across the integrability to chaos crossover for the TFIM (a-c) and XXZ (d-f) respectively. (a,d) Variance of the SDGR \(\sigma_X^2/\Gamma\) as a function of the integrability breaking parameter. (b,e) Skewness of the SDGR \(\gamma_1^X\) as a function of integrability breaking parameter. Vertical dashed lines mark the crossover obtained from \(\| \mathcal A_X \|^2/N\). (c,f) SDGR as a function of the rescaled energy and integrability breaking strength for \(N\!=\!16\) for the TFIM and \(N\!=\!18\) for the XXZ chain. These surfaces extend the representative energy resolved histograms presented in the main text.}
	\label{fig:sm_other_stat_momo_figs}
\end{figure*}

In the main text we motivate the entropy of the SDGR and related quantities as powerful objects which encode the complexity of the geometric sensitivity across the spectral manifold. The statistical moments of the normalized SDGR behave as a witness to the integrable-to-chaotic transition in analogy to the AGP norm, but can further be utilized to identify how much and where in the spectrum the response lives. The second moment:
\begin{equation}
    \sigma^2_{\lambda} = \int d E \left[ E - \langle E \rangle_\lambda\right]^2 P_\lambda(E) = \frac{\sum_n \left[ E_n - \langle E \rangle_\lambda\right]^2 \chi_\lambda^n}{\sum_n \chi_\lambda^n}\;,
    \label{eq:var_dist}
\end{equation}
identifies the spectral width over which the geometric response is distributed. A small variance indicates that the geometric response is dominated by eigenstates concentrated within a narrow window of energy around the geometric center of mass \(\langle E \rangle_\lambda\). A large variance indicates that the geometric weight is carried by eigenstates with widely separated energies. 

The variance in Fig.~\ref{fig:sm_other_stat_momo_figs}(a) and (d) measures the energy range over which the response is spread for the TFIM and XXZ chains respectively in units of the spectral bandwidth \(\Gamma\). It is approximately constant in the integrable regime, but becomes strongly non-monotonic across the crossover. In the TFIM, the entropy decrease is ultimately accompanied by a reduction of the variance, indicating that the response becomes concentrated both in fewer eigenstates and within a narrower energy window. At stronger perturbations, the mean and variance increase again, consistent with resonances spreading into a broader part of the spectrum due to chaotic mixing. The XXZ chain displays a sharper and more size-dependent structure, suggesting that the local defect reorganizes the response less uniformly, with different energy windows becoming resonant at different perturbation strengths.

The skewness of the SDGR,
\begin{equation}
    \label{eq:skew_sdgr}
    \gamma_1^\lambda = \int dE \left[ \frac{E - \langle E \rangle_\lambda}{\sigma_\lambda}\right]^3 P_\lambda(E) = \frac{\sum_n \chi_\lambda^n (E_n - \langle E \rangle_\lambda)^3}{(\sigma_\lambda)^3\sum_m \chi_\lambda^m }\;
\end{equation}
with \(\sigma_\lambda\) the standard deviation of the SDGR, can be interpreted as a measure of the asymmetry in spectral response. Contextually, \(\gamma_1^\lambda\!>\!0\) means the eigenbasis deformation has a tail toward higher energies where most of the geometric weight lies below its own center of mass with a smaller number of sensitive states extending toward the upper spectrum; \(\gamma_1^\lambda\!<\!0\) means that the response tail extends toward lower energies where the main body of response lies at relatively high energies with exceptionally sensitive states extending toward the lower spectrum.

For the physical models studied, the skewness allows us to understand if the initial concentration developed at \(Y^*\) (with \(Y\) the integrability breaking parameter) develops symmetrically around the center or have a preferential development toward lower or higher ends of the spectrum, i.e., it allows us to add directional information complementary to the entropy in the main text. The behavior of the skewness is shown in  Fig.~\ref{fig:sm_other_stat_momo_figs}(b) and (e) for the TFIM and XXZ chains respectively in units the bandwidth \(\Gamma\). For the TFIM we see that \(\gamma_1^\omega\!\simeq\!0\) at the weakly perturbed integrable regime (in the entropic plateau regime in Fig.~\ref{fig:entropy_results}(a) of the main text when \(S_\omega / \log \mathcal D\!\simeq\!1\)). At the crossover scale \(Y^*\) the skewness becomes non-monotonic, with both positive and negative deviations. The behaviour is system size dependent with larger system sizes preferring a positive deviation. This means that the crossover is not simply symmetric narrowing of the SDGR---as seen explicitly in the corresponding histograms of the SDGR in Fig.~\ref{fig:sm_other_stat_momo_figs}(a)(ii). Instead, different sides of the spectrum dominate sequentially and the direction of the anomalous tail changes during the spectral reorganization. The intermediate behaviour between \(Y^*\) and strong integrability breaking is complex, but for sufficiently strong perturbation strengths \(Y\!\gg\!Y^*\) the skewness becomes strongly negative and then suddenly resurges to a positive value; indicating the geometric sensitivity bounces between highly sensitive states in both low and high energy parts of the spectrum. 

For the XXZ model [panel (e)] the skewness is constant and negative during the weak integrable plateau. The curves do not converge to an unique value like in the TFIM, although as the system size gets larger there is a tendency toward zero. We attribute this to the presence of the local site defect. At the onset scale \(Y^*\) the skewness becomes strongly negative indicating a persistent tail of sensitivity is concentrated toward lower energies that is amplified during the crossover regime. Interestingly the characteristic behaviour is consistent for different system sizes.  

In Fig.~\ref{fig:sm_other_stat_momo_figs}(c) and (f) we show the evolution of the SDGR across the full integrability to chaos crossover from the TFIM (\(N\!=\!16\)) and XXZ (\(N\!=\!18\)) chains respectively. The histograms in Fig.~\ref{fig:summary_stat}(a) and (b) correspond to snapshot of this surface plot at specific values of integrability parameter \(Y\!=\!(h_x,\epsilon_d)\). The horizontal axis gives the position within the rescaled many-body spectrum, the second axis gives the perturbation strength on a logarithmic scale, and the surface height represents the geometric response carried by each spectral region. Despite the different microscopic mechanisms for integrability breaking, the evolution of the geometric deformation across the spectrum supports the interpretation in the main text. In the weakly perturbed integrable regime, the geometric weight is relatively smoothly spread throughout the spectrum. Near \(Y^*\), the smooth response becomes uneven and concentrates within selected windows of energy and the onset of chaos is not homogeneous across the eigenbasis. It begins through spectrally local rearrangements in which a restricted set of eigenstates become exceptionally sensitive before the response spreads across the rest of the spectrum. The surface plots also provide insight into the physical meaning of the entropic minimum. At \(Y^{\text{min}}\), the eigenbasis deformation becomes localized and the most pronounced concentration
of the geometric weight occurs at selected spectral regions. For stronger perturbations, these regions broaden and the progression follows spectrally local nucleation followed by spectral spreading.

\section{Finite temperature extension}\label{sm_finite_temp_ext}

The SDGR admits a natural extension to finite temperature through inclusion of the Boltzmann factor of each eigenstate:
\begin{equation}
    \label{sm_sdgr_finite_temp}
    B_\lambda(E,\beta) = \sum_n \chi_\lambda^n e^{-\beta E_n} \delta(E-E_n)\;,
\end{equation}
where the original SDGR Eq.~\eqref{eq:fid_suc_dos} is recovered at the infinite temperature limit \(\beta\!=\!0\). In this construction one may write the geometric partition function Eq.~\eqref{eq:sm_geo_pf} as
\begin{equation}
    \mathcal Z_\lambda^\chi(\beta) = \int dE B_\lambda(E,\beta) = \sum_n \chi_\lambda^n e^{-\beta E_n}\;,
\end{equation}
which can equivalently be written in terms of the AGP (in the parallel transport gauge) as
\begin{equation}
    \mathcal Z_\lambda^\chi(\beta) = \text{Tr}\left[e^{-\beta \hat H} \mathcal A_\lambda^2 \right]\;.
\end{equation}
The ordinary canonical partition function \(Z(\beta)\) can therefore be related to the geometric thermal partition function via
\begin{align}
    \mathcal Z_\lambda^\chi(\beta) &= Z(\beta) \text{Tr}[\hat \varrho_\beta \mathcal A_\lambda^2] \\
    &= Z(\beta) \sum_n \frac{e^{-\beta E_n}}{Z(\beta)} \chi_\lambda^n\;.
\end{align}
If we consider the ordinary canonical average \(\langle X \rangle_\beta \!\equiv\![Z(\beta)]^{-1}\sum_n X_n e^{-\beta E_n}\) we can show that the geometric partition function can be obtained as
\begin{equation}
    \mathcal Z_\lambda^\chi(\beta) = Z(\beta) \langle \chi_\lambda\rangle_\beta\;,
\end{equation}
i.e., \(\langle \chi_\lambda\rangle_\beta = \mathcal Z_\lambda^\chi(\beta)  / Z(\beta)\) and thus, the geometric partition function is the ordinary partition function multiplied by the thermal average fidelity susceptibility. Interestingly we can also show that
\begin{equation}
    \mathcal Z_\lambda^\chi(\beta) = Z(\beta)\|\mathcal A_\lambda\|^2_\beta\;,
\end{equation}
which defines the finite temperature AGP fluctuations in the presence of thermal fluctuations. This provides a compact formalism for interrogating the competition between eigenbasis deformations in the presence of thermal fluctuations. For the normalized finite temperature SDGR, the corresponding probability assigned to each eigenstate \(\vert n \rangle\) is therefore
\begin{equation}
    p_{\lambda}^n(\beta) = [\mathcal Z_\lambda^\chi(\beta)]^{-1}\chi_\lambda^n e^{-\beta E_n}\;,
\end{equation}
and the normalized energy resolved distribution therefore follows as 
\begin{equation}
    P_\lambda(E,\beta) = [\mathcal{Z}_\lambda^\chi(\beta)]^{-1} B_\lambda(E,\beta) = \sum_n p^n_\lambda(\beta) \delta(E-E_n)\;.
\end{equation}
In this formulation the Shannon entropy can be written in the rather suggestive form:
\begin{equation}
    S_\lambda(\beta) = \log \mathcal Z_\lambda^\chi(\beta) + \beta \langle E \rangle_{\lambda,\beta} - \langle \log \chi^n_\lambda \rangle_{\lambda,\beta}
\end{equation}
[where \(\langle E \rangle_{\lambda,\beta}\!\equiv\!\sum_n p_\lambda^n(\beta) E_n\)] which takes the form of a generalized thermodynamic identity, \(S = \log Z + \beta \langle E \rangle\), with an additional contribution arising from the state dependent geometric weight, given by the ensemble averaged logarithmic fidelity susceptibility. The geometric contribution to the entropy admits an equivalent information theoretic interpretation. To see this explicitly, we compare the geometrically weighted probabilities \(p_\lambda^n(\beta)\) with the canonical probabilities \(q_n(\beta) = [Z(\beta)]^{-1}e^{-\beta E_n}\). The Kullback-Leibler divergence is given by
\begin{align}
    D_{KL}(p_\lambda \| q) &= \sum_n p_\lambda^n(\beta) \log{\frac{p_\lambda^n(\beta)}{q_n(\beta)}} \\
    &= \langle \log \chi_\lambda^n \rangle_{\lambda,\beta} + \log{Z(\beta)} - \log{\mathcal Z_\lambda^\chi(\beta)}\;.
\end{align}
Using \(\mathcal Z_\lambda^\chi(\beta) \!=\!Z(\beta)\langle \chi_\lambda\rangle_\beta\) this can be expressed as
\begin{equation}
    D_{KL}(p_\lambda \| q) = \langle \log \chi_\lambda^n \rangle_{\lambda,\beta} - \log \langle \chi_\lambda \rangle _\beta
\end{equation}
and substituting this back into the entropy finally gives
\begin{equation}
    S_\lambda(\beta) = \log Z(\beta) + \beta \langle E \rangle_{\lambda,\beta} - D_{KL}(p_\lambda\|q)
\end{equation}
and the entropy of the finite temperature SDGR can be written in a thermodynamic form involving the ordinary partition function with a correction determined by the statistical distinguishability between the geometrically weighted and canonical eigenstate distributions. If we define the conventional canonical entropy as
\begin{equation}
    S_{th}(\beta) = \log Z(\beta) + \beta \langle E \rangle_\beta
\end{equation}
we can obtain
\begin{equation}
    S_\lambda(\beta) = S_{th}(\beta) + \beta[\langle E \rangle_{\lambda,\beta} - \langle E \rangle_\beta] - D_{KL}(p_\lambda\|q)\;.
\end{equation}

The SDGR admits a natural statistical interpretation. Its total spectral weight defines a geometric partition function
\begin{equation}
    \mathcal Z_\lambda^\chi = \int dE B_\lambda(E) = \sum_n \chi_\lambda^n = \mathcal D \| \mathcal A_\lambda \|^2\;,
\end{equation}
which measures the total deformation of the eigenbasis generated by a variation in \(\lambda\). The normalized weight \(p_\lambda^n = \chi_\lambda^n / \mathcal Z_\lambda^\chi\) is the fraction of the total response carried by the eigenstate \(\vert n \rangle\). The normalized SDGR can therefore be written as 
\begin{equation}
    P_\lambda(E) = [\mathcal{Z}_\lambda^\chi]^{-1} B_\lambda(E) = \sum_n p_\lambda^n \delta(E-E_n)\;,
\end{equation}
describing the energy distribution obtained by sampling eigenstates according to their geometric sensitivity. The entropy of the geometric ensemble then takes the form
\begin{equation}
    S_\lambda = \log \mathcal Z_\lambda^\chi - \langle \log \chi_\lambda^n \rangle_{p_\lambda}\;,
\end{equation}
where \(\langle \log \chi_\lambda^n \rangle_{p_\lambda} = \sum_n p_\lambda^n \log \chi_\lambda^n\). Equivalently the entropy can be written in terms of the typical fidelity susceptibility~\cite{LeBlond_2021} as \(S_\lambda = \log{(\mathcal Z_\lambda^\chi / \chi_\lambda^{\text{typ}})}\) and therefore the entropy compares the total geometric response with the typical response encountered when eigenstates are sampled according to their fidelity susceptibility weight. 

The geometric partition function additionally acts as a cumulant generating function for the energy distribution of the finite temperature SDGR. In particular, we define
\begin{equation}
    \Phi_\lambda(\beta) = \log \mathcal Z_\lambda^\chi(\beta)\;,
\end{equation}
where successive derivatives with respect to inverse temperature \(\beta\) generate energy cumulants of the geometrically weighted distribution
\begin{equation}
    \kappa_{r}^\lambda(\beta) = (-1)^r \frac{\partial^r}{\partial \beta^r}\log \mathcal Z_\lambda^\chi(\beta)\;,
\end{equation}
for example the first derivative yields the mean energy
\begin{equation}
    -\partial_\beta \log \mathcal Z_\lambda^{\chi}(\beta) = \langle E \rangle_{\lambda,\beta}\;.
\end{equation}
Because \(\mathcal Z_\lambda^\chi(\beta) = Z(\beta)\langle \chi_\lambda \rangle_\beta\) these cumulants can be decomposed equivalently as
\begin{equation}
    \kappa_r^\lambda(\beta) = (-1)^r \frac{\partial^r}{\partial \beta^r} \log Z(\beta) + (-1)^r \frac{\partial^r}{\partial \beta^r}\log \langle \chi_\lambda \rangle_\beta\;,
\end{equation}
showing the finite temperature SDGR encodes modifications of ordinary thermal energy fluctuations induced by geometric reweighting of the eigenbasis. 


\section{Connection to linear response}\label{sm_linear_reponse}
The finite temperature geometric partition function admits a direct connection to linear response theory. We define the perturbation operator as \(\hat V = \partial_\lambda \hat H\) and for consistency with the main text, we present the result for the regularized fidelity susceptibility:
\begin{equation}
    \chi_\lambda^n = \sum_{m\neq n} \frac{\omega^2_{mn}}{(\omega_{mn}^2 + \mu^2)^2} \vert \langle m \vert \hat V _\lambda\vert n \rangle \vert^2\;,
\end{equation}
with \(\omega_{mn}\!=\!E_m - E_n\). Introducing the spectral function:
\begin{equation}
    C_\lambda(\omega,\beta) = [Z(\beta)]^{-1} \sum_{n,m\neq n} e^{-\beta E_n} \vert \langle m \vert \hat V_\lambda \vert n \rangle \vert^2 \delta(\omega - \omega_{mn})\;,
\end{equation}
the thermally averaged fidelity susceptibility becomes
\begin{equation}
    \frac{\mathcal Z_\lambda^\chi(\beta)}{Z(\beta)} = \langle \chi_\lambda \rangle_\beta = \int_{-\infty}^{\infty} d\omega \frac{\omega^2}{(\omega^2 + \mu^2)^2}C_\lambda(\omega,\beta)\;.
\end{equation}
Through the fluctuation dissipation theorem, the spectral function is related to the dissipative part of the Kubo response through
\begin{equation}
    \mathcal R_\lambda''(\omega,\beta) = \pi(1 - e^{-\beta \omega)}C_\lambda(\omega,\beta)\;,
\end{equation}
provided \(\omega \!>\!0\). Detailed balance gives
\begin{equation}
    C_\lambda(-\omega,\beta) = e^{-\beta\omega}C_\lambda(\omega,\beta)\;.
\end{equation}
From the above we can therefore arrive at the following relationship:
\begin{equation}
    \frac{\mathcal Z_\lambda^\chi(\beta)}{Z(\beta)} =   \frac{1}{\pi}\int_0^\infty d\omega \frac{\omega^2}{(\omega^2+\mu^2)^2} \coth{\left(\frac{\beta \omega}{2}\right)} \mathcal R''_\omega(\omega,\beta)\;.
\end{equation}
Remarkably, the finite temperature geometric partition function is determined by a frequency weighted moment of the dissipative response associated with the perturbation that generates the deformation of the eigenbasis. In the unregularized limit, \(\mu\!\to\!0\), the kernel reduces to the usual \(1/\omega^2\), recovering the familiar interpretation of fidelity susceptibly as an inverse second frequency moment of the dynamical response function~\cite{PhysRevE.76.022101,zoller,PhysRevLett.105.120501}.

\section{Alternative constructions for Random Matrix Theory}\label{sm_rmt}

Inspired by the construction in Ref. \cite{lenz1991}, we consider a model of random matrices with the following structure
\begin{equation}
    H(\lambda,\beta) = \mathcal{N}(\beta)(H_0 + \lambda H_1 + \beta  V)
    \label{eq:rmt_hami_sm}
\end{equation}
where the normalization is $\mathcal{N}(\beta)^2 = (1+\lambda_0^2+\beta^2)^{-1}$ is a normalization factor. The parameter $\beta$ drives a transition from integrability to chaos. We take $\lambda$ to be the parameter with respect to which the derivatives of the eigenstates are taken in the definition of the fidelity susceptibility. The key point of this construction is choosing $H_0$ and $H_1$ in such a way that $H(\lambda,0)$ is integrable for all $\lambda$. We do this in two ways. First, for the Pauli-RMT model described in the main text, we have
\begin{align}
    \hat H_0 &= \sum\limits_{k=1}^N c_k^{(0)} \hat \sigma^z_k \\
    \hat H_1 &= \sum\limits_{k=1}^N (c_k^{(1)}\hat \sigma^z_k+d_k^{(1)}\hat \sigma^x_k)\\
    \hat V &= \sum\limits_{k=1}^N (c_k^{(2)}\hat \sigma^z_k+d_k^{(2)}\hat \sigma^x_k) + \sum\limits_{k\neq l=1}^N\sum\limits_{p_1,p_2=x,z} f_{kl}^{p_1,p_2}\hat \sigma_k^{p_1} \hat \sigma_l^{p_2}.
\end{align}
Note that $\hat \sigma_y$ operators are avoided to keep $H$ real. In the Pauli-RMT construction, $\hat H_0+\lambda \hat H_1$ is trivially integrable as it corresponds to noninteracting particles. The interaction term $V$ thus drives a transition to chaos. The coefficients in the three operators are chosen randomly from a uniform distribution in the range $[-1,1]$; each set of parameters is also normalized to $\sum_l g_l^2=1$, where $g$ here denotes any of $c$, $d$, or $f$. 

A more standard random matrix construction is as follows \cite{lenz1991}. We take $\hat H_0$ to be a diagonal matrix whose entries are uniformly distributed in the interval $[-\pi/2,\pi/2]$ and $\hat V$ to be a matrix drawn from the Gaussian Orthogonal Ensemble (namely, a symmetric matrix where each entry is a normally distributed with mean zero). To ensure the integrability of $\hat H(\lambda,0)$, we consider $\hat H_1$ to be block diagonal, containing $k_B$ blocks, each of them being themselves GOE matrices. We call this model GOE-RMT. This construction depends on the number \(k_B\) of blocks considered. We point out that alternative constructions of families of integrable random matrices exist \cite{scaramazza2016}, but we have not explored them.

For completeness, we recall that to characterise the degree of quantum chaos, the standard metric is the mean level spacing ratio (MLSR) $\langle r \rangle$, which captures the emergence of level repulsion in the bulk of the spectrum~\cite{Atas2013}. For ordered eigenvalues $\{e_i\}$, adjacent level spacings are defined as $\Delta_j = e_{j+1}-e_j$ 
and successive spacings are compared through \begin{equation}
r_j = \frac{\min(\Delta_j,\Delta_{j+1})} {\max(\Delta_j,\Delta_{j+1})}.
\end{equation} The MLSR is then obtained by averaging $r_j$ over the spectrum (in a RMT setting). The RMT-GOE prediction for the MLSR is $\overline{r}_{\rm GOE}\simeq 0.535$~\cite{Atas2013}. By contrast, integrable systems exhibit Poissonian level statistics with uncorrelated eigenvalues, giving $\overline{r}_{\rm INT}\simeq 0.386$~\cite{Atas2013}.

\begin{figure}
    \centering
    \includegraphics[width=1\linewidth]{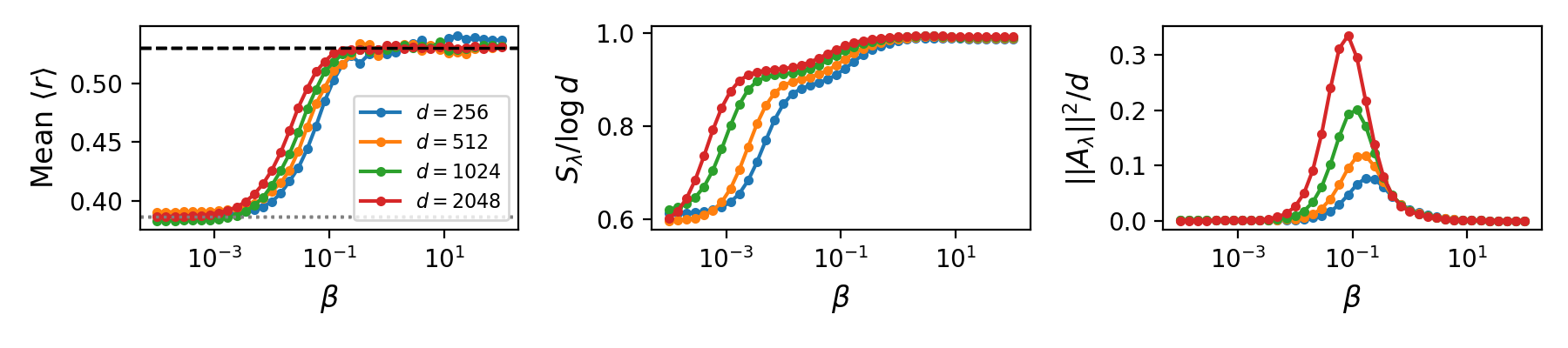}
    \includegraphics[width=1\linewidth]{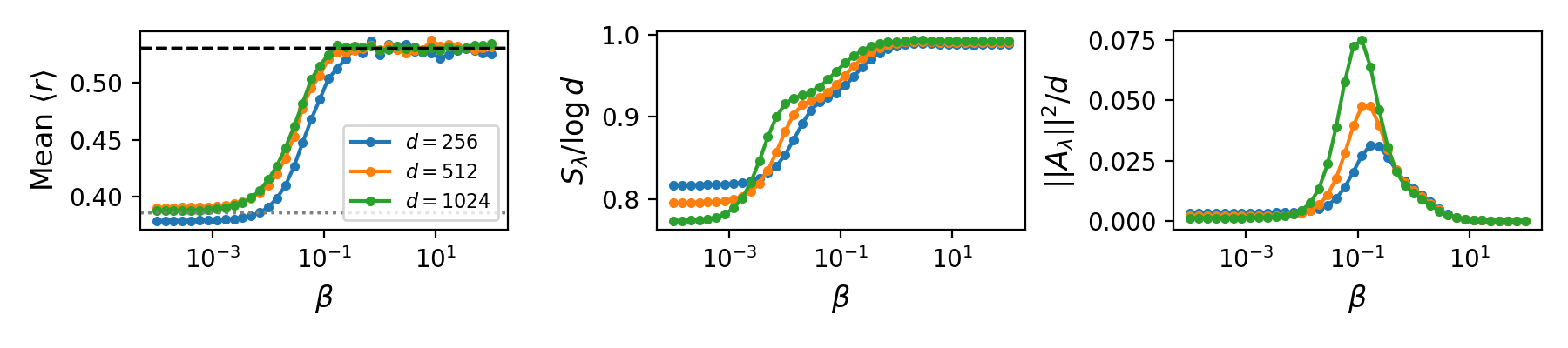}
    \caption{Panels show (left) mean level spacing ratio, (middle) entropy of the SDGR $S_\lambda$ and (right) norm of the AGP $||\mathcal{A}_\lambda||$ as a function of the integrability breaking parameter $\beta$ for the RMT-GOE model described in this section. Top figure corresponds to $k_B=d/2$, bottom figure to $k_B=d/8$, illustrating two different ways of constructing the model.}
    \label{fig:sm_rmt_goe}
\end{figure}

\begin{figure}
    \centering
    \includegraphics[width=1\linewidth]{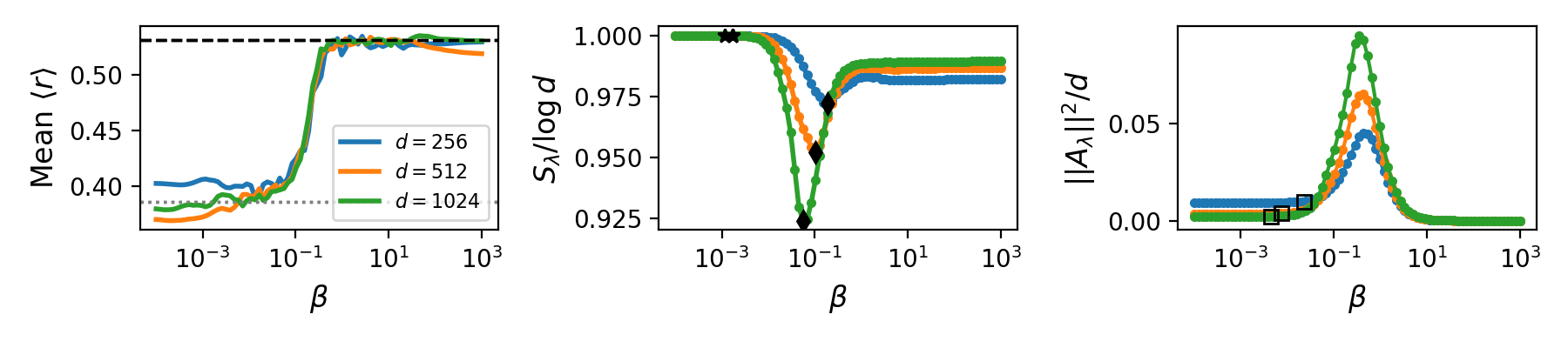}
    \includegraphics[width=1\linewidth]{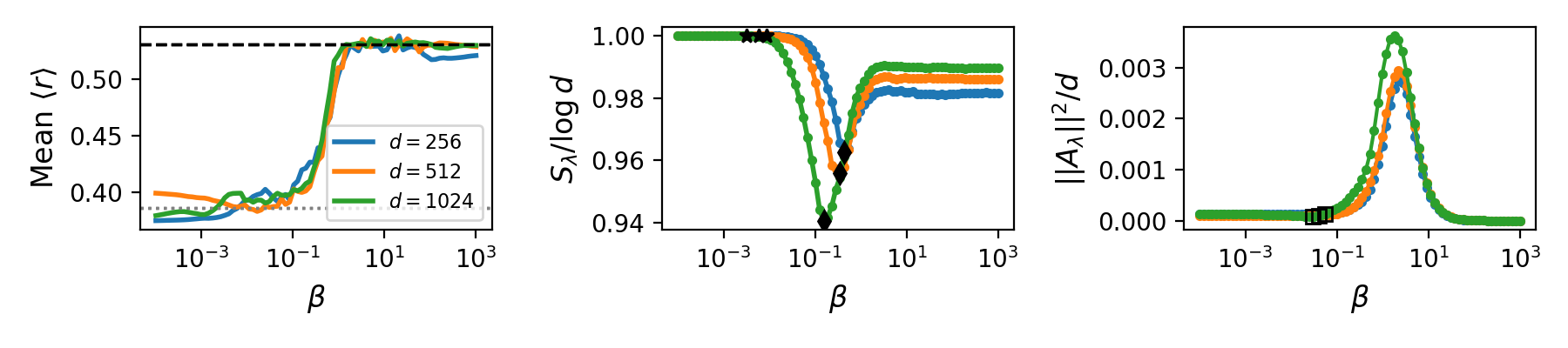}
    \caption{Panels show (left) mean level spacing ratio, (middle) entropy of the SDGR $S_\lambda$ and (right) norm of the AGP $||\mathcal{A}_\lambda||$ as a function of the integrability breaking parameter $\beta$ for the RMT-Pauli model. Top figure corresponds to $\lambda_0=0.5$, bottom figure to $\lambda_0=3$, and note figure in main text is $\lambda_0=1$.}
    \label{fig:sm_rmt_pauli}
\end{figure}

Here we present results for both RMT constructions to complement Fig.~\ref{fig:RMT} in the main text. In Fig. \ref{fig:sm_rmt_goe} we plot the mean level spacing ratio, together with the entropy of the SDGR $S_\lambda$ and the norm of the AGP $||\mathcal{A}_\lambda||$ for the GOE-RMT construction. The geometric response is evaluated around $\lambda=\lambda_0=1$, but we have checked that the overall results do not depend on this choice significantly. All quantities are shown as a function of the integrability-breaking parameter $\beta$ and for different system sizes $d=\log_2(N)$. As expected we find that the GOE-RMT model leads to a integrability-chaos transition, here around $\beta\sim 0.1$, signaled by the crossover of $\langle r\rangle$. The AGP norm correspondingly shows an exponential in $N$ increase at this point. However we do not find the characteristic dip in the SDGR entropy, and actually $S_{\lambda}$ does not display any prominent feature at the chaos onset. As seen from the figure, this is true for different choices of $k_B$, the number of blocks in $H_1$ in the RMT-GOE construction. These results are in contrast with the case of Pauli-RMT, where the dip is indeed observed. We complement the results in the main text with the plots in Fig. \ref{fig:sm_rmt_pauli}, displaying similar behavior for other choices of $\lambda_0$.

\end{document}